\documentclass{article}
\usepackage{hyperref}
\usepackage{xurl}
\usepackage{datetime}
\usepackage{graphicx}
\usepackage{natbib}
\usepackage{amsmath}
\usepackage{booktabs}

\hypersetup{colorlinks=true, allcolors=blue}

\title{A Semantic Approach to the Academic Publishing Network: Document Vector Representations and Hybrid Structural–Semantic Fusion over OpenAlex Data}

\author{Robert Šamárek \and Radek Martinek}

\newdate{articleDate}{13}{6}{2026}
\date{\displaydate{articleDate}}

\begin{document}
\maketitle
\begin{abstract}Structural graph analysis of the academic publishing network captures the topological relationships between entities but does not see the content of works. Building on our structural approach, this work complements it with a semantic layer and a parameterized structural--semantic fusion. We represent scientific documents by citation-informed vector embeddings (SPECTER2) and store them in an embedded vector database keyed by the stable OpenAlex ID, so that they connect directly to the graph layer. We define a modular late-fusion function that combines semantic similarity (cosine of embeddings) and structural similarity (bibliographic coupling) with a tunable weight $\alpha$ whose value is chosen according to the specific task. On the corpus of VSB -- Technical University of Ostrava we show two things: citation-informed embeddings agree with the expert OpenAlex topical taxonomy better than a TF-IDF baseline, and in a recommendation use case the structural, semantic, and combined signals carry information in different regimes depending on the available data. Hybrid fusion here is not a universally better method but an explicit mechanism for steering complementary signals according to the task. We release the whole approach as an open-source extension of the apnet library with a reproducible workflow.\end{abstract}

\section{Introduction}

The graph analysis of the academic publishing network that we presented in the previous work [Šamárek \& Martinek, in preparation] captures the \textbf{topology} of relationships between works, authors, institutions, and topics, and enables community detection, the computation of centralities, and the screening of anomalous publishing patterns. The structural view has, however, one principled limitation: it \textbf{does not see the content} of works. Two works on the same problem that neither cite each other nor share collaborators remain distant in a purely structural model, even though they are close in content. Moreover, the structural signal depends on a sufficient citation and collaboration trace, which for some works is sparse or missing; a content representation is a complementary source of information in such cases.

In parallel with the development of graph analytics, a revolution took place in text representation. The line of development from \textbf{SciBERT} \citep{beltagySciBERTPretrainedLanguage2019} through \textbf{SPECTER} \citep{cohanSPECTERDocumentlevelRepresentation2020} and \textbf{SciNCL} \citep{ostendorffNeighborhoodContrastiveLearning2022} to \textbf{SPECTER2} \citep{singhSciRepEvalMultiFormat2022} produced dense vector representations of scientific documents in which content-similar works lie close together. These representations are complementary to the graph topology: the graph captures \textit{who works with whom and what cites what}, the embeddings capture \textit{what a work is about}.

The present work builds on the structural approach and complements it with a \textbf{semantic layer} and a \textbf{hybrid fusion}. Our contributions are three:

\begin{enumerate}
\item \textbf{A semantic layer.} We represent scientific works by citation-informed embeddings (SPECTER2) over the title and abstract and store them in an embedded vector database keyed by the stable OpenAlex ID; the layer thus connects directly to the graph model without remapping.
\item \textbf{A hybrid fusion layer.} We propose a modular late-fusion mechanism that combines semantic similarity (cosine of embeddings) and structural similarity (bibliographic coupling) with a tunable weight, and we apply it to the recommendation of related works.
\item \textbf{An empirical demonstration.} On an institutional corpus we show that (i) citation-informed embeddings agree with the expert OpenAlex taxonomy better than TF-IDF, and (ii) in a recommendation use case the structural, semantic, and combined signals carry information in different regimes---fusion is a parameterized mechanism chosen according to the task, not a universally winning method. All as a reproducible open-source extension of the apnet library.
\end{enumerate}

\section{Related work}

\subsection{Vector representations of scientific text}

Contextual language models of the BERT type \citep{devlinBERTPretraining2019} gave rise to a branch of domain-specific models for science: \textbf{SciBERT} \citep{beltagySciBERTPretrainedLanguage2019} and especially models trained on the citation signal---\textbf{SPECTER} \citep{cohanSPECTERDocumentlevelRepresentation2020} uses triplets (citing--cited--non-cited) to learn representations in which citation proximity is encoded directly into the geometry of the space; \textbf{SciNCL} \citep{ostendorffNeighborhoodContrastiveLearning2022} extends this with contrastive learning in the citation neighbourhood, and \textbf{SPECTER2} \citep{singhSciRepEvalMultiFormat2022} adds task-specific adapters and achieved state-of-the-art performance on the \textbf{SciRepEval} benchmark at the time of its publication. For generating sentence/document embeddings in general, \textbf{Sentence-BERT} \citep{reimersSentenceBERT2019} and contrastive approaches such as \textbf{SimCSE} \citep{gaoSimCSE2021} became established; modern universal embedders (\textbf{E5} \citep{wangE5TextEmbeddings2024}, \textbf{BGE} \citep{xiaoCPackBGE2024}, \textbf{GTE} \citep{liGTETextEmbeddings2023}, \textbf{NV-Embed} \citep{leeNVEmbed2024}) push the boundary further. As a classical sparse baseline we use \textbf{TF-IDF} within the probabilistic relevance framework \citep{robertsonProbabilisticRelevanceFramework2009}.

\subsection{Vector databases and approximate search}

Nearest-neighbour search in a high-dimensional space is addressed by libraries and databases for approximate search (ANN). \textbf{FAISS} \citep{douzeFaissLibrary2024, johnsonBillionScaleGPU2021} offers scalable indexes including \textbf{product quantization} \citep{jegouProductQuantization2011}, the graph index \textbf{HNSW} \citep{malkovEfficientRobustHNSW2020} provides an excellent speed/accuracy trade-off, and \textbf{ScaNN} \citep{guoAnisotropicVectorQuantization2020} introduces anisotropic quantization. Embedded vector databases such as \textbf{LanceDB} \citep{lancedb2024} and server systems such as \textbf{Qdrant} \citep{qdrant2024} make these algorithms available at the application level; thanks to its in-process deployment and columnar format, LanceDB naturally complements the graph layer without server infrastructure.

\subsection{Hybrid search and recommendation}

Dense (semantic) and sparse (lexical) or structural signals are complementary \citep{luanSparseDenseAttentional2021}. Hybrid search combines them---from \textbf{reciprocal rank fusion} \citep{cormackReciprocalRankFusion2009} through \textbf{DPR} \citep{karpukhinDensePassageRetrieval2020}, \textbf{ColBERT} \citep{khattabColBERT2020}, to \textbf{SPLADE} \citep{formalSPLADE2021}. In scientometrics, the recommendation of scientific literature has a rich tradition \citep{beelResearchPaperRecommender2016, baiScientificPaperRecommendation2019, mcneeRecommendingCitations2002}; a structural signal can be drawn from graph embeddings (\textbf{node2vec} \citep{groverNode2vec2016}, \textbf{GraphSAGE} \citep{hamiltonGraphSAGE2017}) or classical link prediction \citep{libennowellLinkPrediction2007}. Hybrid combinations of graph and content have proven beneficial for recommendation \citep{liuAcademicLiteratureRecommendation2025, kanwalResearchPaperRecommendation2024}; the \textbf{Graph-RAG} paradigm \citep{lewisRetrievalAugmentedGeneration2020, edgeGraphRAG2024, sarmahHybridRAGIntegrating2024} connects these signals with generative models (beyond the scope of this work). The structural and content signals are complementary also in that content carries information even where the citation signal of a new work is missing or sparse \citep{scheinColdStartRecommendations2002}.

\subsection{Evaluation}

We measure ranking quality with the metrics \textbf{NDCG} \citep{jarvelinCumulatedGain2002}, \textbf{MAP}, and \textbf{MRR}; the agreement of clustering with a reference taxonomy with \textbf{NMI} \citep{strehlGhoshClusterEnsembles2002} and \textbf{ARI} \citep{hubertArabieComparingPartitions1985}.

\section{Semantic model}\label{sec-sem-model}

We represent each work $w$ by a vector $\mathbf{e}_w \in \mathbf{R}^{768}$ obtained from the \textbf{SPECTER2} model \citep{singhSciRepEvalMultiFormat2022} from a concatenation of the title and abstract. SPECTER2 builds on \texttt{allenai/specter2\_base} (a BERT architecture) with a \textbf{proximity adapter} trained on the citation signal---the resulting embeddings therefore carry not only lexical but also citation-thematic proximity. We reconstruct the abstract from the \texttt{abstract\_inverted\_index} field of OpenAlex; works without an abstract we represent by the title alone. We L2-normalize the vectors, so that similarity is given by the cosine.

We store the embeddings in the embedded vector database \textbf{LanceDB} \citep{lancedb2024} keyed by the stable OpenAlex ID, so that each representation is directly linked to the corresponding node of the graph layer. As a classical baseline we use \textbf{TF-IDF} (with dimensionality reduction via TruncatedSVD) in order to separate the contribution of the citation-informed model from mere lexical overlap.

\section{Hybrid fusion layer}\label{sec-fusion}

The structural and semantic layers produce complementary signals. For a query work $q$ and a candidate $c$ we define:

\begin{itemize}
\item \textbf{semantic similarity} $s_\text{sem}(q,c) = \cos(\mathbf{e}_q, \mathbf{e}_c)$;
\item \textbf{structural similarity} $s_\text{str}(q,c) = J(R_q, R_c)$ as the Jaccard similarity of the reference sets $R_q, R_c$ (bibliographic coupling---two works are structurally close if they share cited sources).
\end{itemize}

We min--max normalize both components to a comparable scale and combine them with \textbf{late fusion} and a weight $\alpha \in [0,1]$:

\begin{equation}
s_\text{hyb}(q,c) = \alpha \, \hat{s}_\text{sem}(q,c) + (1-\alpha)\, \hat{s}_\text{str}(q,c).
\end{equation}

The extreme values $\alpha = 1$ and $\alpha = 0$ correspond to the purely semantic and purely structural method, respectively; intermediate values fuse them. The choice of $\alpha$ is tunable per task---in line with the thesis that the hybrid approach is not a dogma but a mechanism chosen according to the nature of the task and the data.

\section{Implementation}

We implement the semantic and fusion layers as an extension of the open-source library \textbf{apnet} (modules \texttt{semantic.py}, \texttt{hybrid.py}; MIT licence). SPECTER2 is loaded via the \texttt{adapters} library on top of \texttt{transformers}; embedding proceeds in batches on the CPU (on a commodity laptop, on the order of tens of minutes for thousands of works, with the option of GPU acceleration). Heavy dependencies are optional---the apnet core (the graph layer) works without them. The whole experiment is reproducible: a script fetches the abstracts, computes the embeddings, builds the vector index, and runs experiments H2 and H3 from a single source.

\section{Experiments}\label{sec-experimenty}

\subsection{Corpus}

We run the experiments on the institutional corpus of works of VSB -- Technical University of Ostrava (2020--2025), identical to that of our structural approach, which enables a \textbf{direct connection} of the two layers. Of the 7,317 works, 81.6\% have a non-empty abstract; since works without an abstract are represented by their title alone, embeddings could be computed for 7,316 works with available textual metadata and an OpenAlex subfield annotation, divided into 197 subfields.

\subsection{Semantic coherence: agreement of embeddings with the expert taxonomy}

\begin{figure}[!htbp]
\centering
\includegraphics[width=0.75\linewidth]{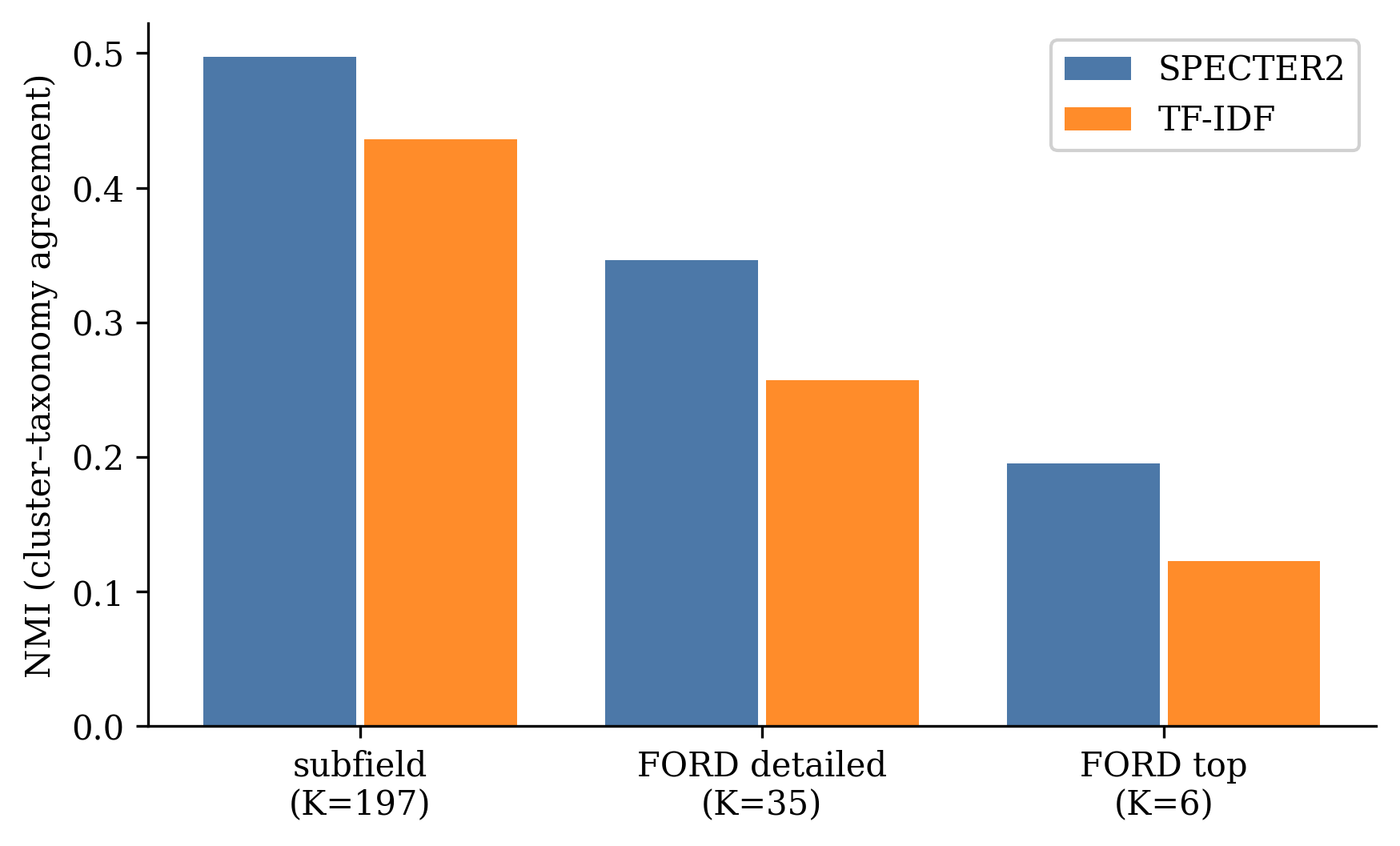}
\caption[]{NMI of embedding clustering against the taxonomy at three granularities (OpenAlex subfield $\rightarrow$ OECD FORD detailed $\rightarrow$ 6 main FORD fields). SPECTER2 outperforms TF-IDF at all granularities.}
\label{fig-h2-clustering}
\end{figure}

\begin{figure}[!htbp]
\centering
\includegraphics[width=0.8\linewidth]{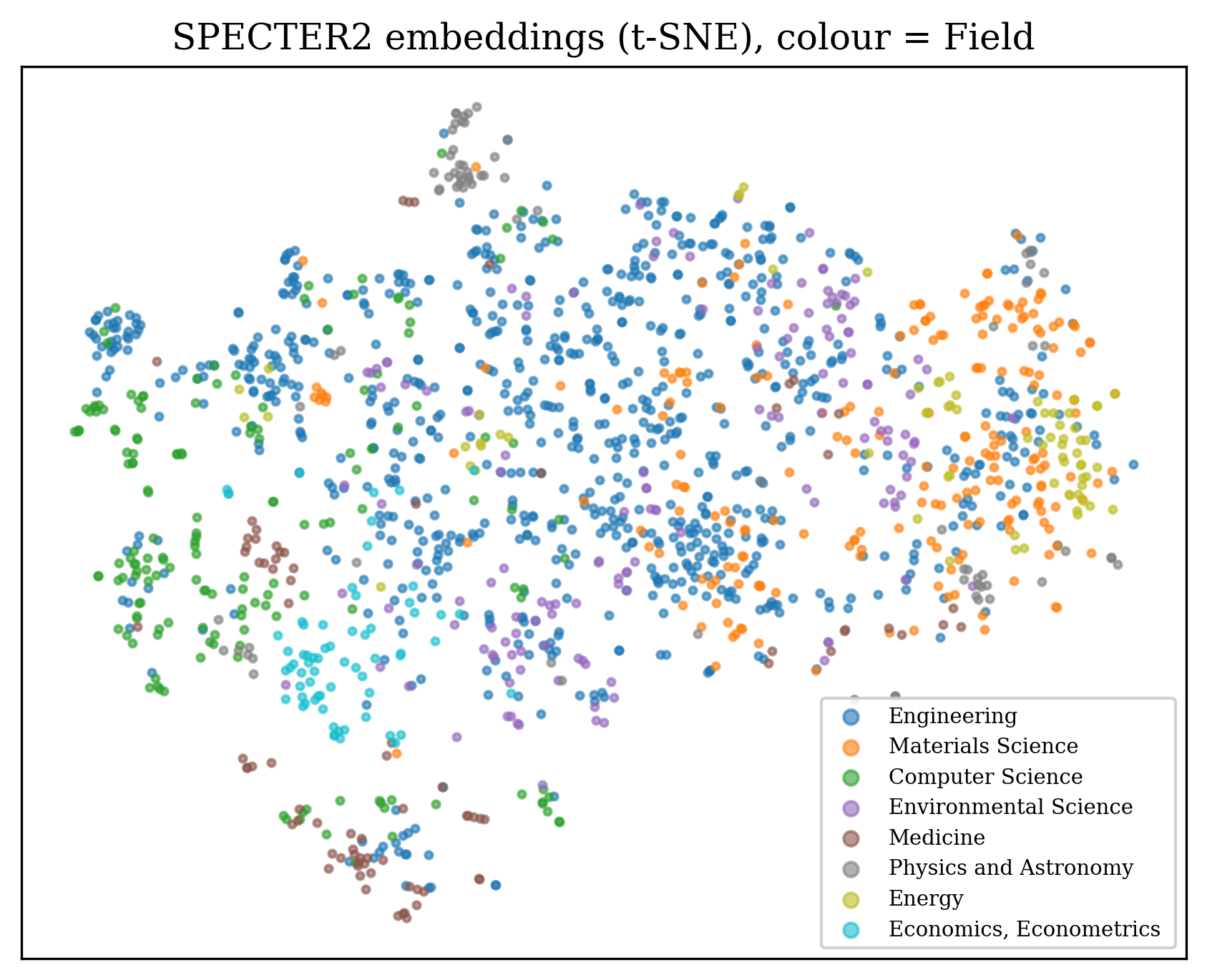}
\caption[]{A two-dimensional t-SNE projection of SPECTER2 embeddings (sample of works), colour corresponds to the field (Field) per OpenAlex. Thematically related works form coherent clusters.}
\label{fig-h2-tsne}
\end{figure}

Citation-informed embeddings should correspond to expertly defined thematic categories better than a lexical baseline. We cluster the embeddings (SPECTER2 and TF-IDF) with the $k$-means algorithm and measure the agreement with the expert taxonomy by NMI and ARI at \textbf{three granularities}: fine OpenAlex subfields (197), coarser \textbf{OECD FORD} detailed fields, and 6 main FORD fields (Table~\ref{tab-h2}, Figure~\ref{fig-h2-clustering}). \textbf{SPECTER2 outperforms TF-IDF consistently across all granularities}---including the internationally standardized OECD FORD taxonomy, where its relative advantage is even more pronounced (NMI 0.195 vs. 0.123). This confirms the robustness of the conclusion; the absolute NMI with the coarser taxonomy decreases, because six broad FORD fields naturally do not separate the clusters as sharply as fine subfields. Figure~\ref{fig-h2-tsne} shows a 2D projection (t-SNE) of the SPECTER2 embeddings coloured by field---the colour clusters document that the geometry of the space corresponds to the thematic structure.

\begin{table}
\centering
\caption[]{Agreement of embedding clustering with the taxonomy at three granularities (subfield $\rightarrow$ OECD FORD; higher = better)}
\label{tab-h2}
\begin{tabular}{p{\dimexpr 0.167\linewidth-2\tabcolsep}p{\dimexpr 0.167\linewidth-2\tabcolsep}p{\dimexpr 0.167\linewidth-2\tabcolsep}p{\dimexpr 0.167\linewidth-2\tabcolsep}p{\dimexpr 0.167\linewidth-2\tabcolsep}p{\dimexpr 0.167\linewidth-2\tabcolsep}}
\toprule
Taxonomy & K & SPECTER2 NMI & SPECTER2 ARI & TF-IDF NMI & TF-IDF ARI \\
\hline
OpenAlex subfield & 197 & 0.497 & 0.077 & 0.436 & 0.053 \\
FORD detailed (OECD) & 35 & 0.346 & 0.135 & 0.257 & 0.082 \\
FORD main (OECD) & 6 & 0.195 & 0.123 & 0.123 & 0.055 \\
\bottomrule
\end{tabular}
\end{table}

\subsection{Use case: recommendation and signal fusion}

\begin{figure}[!htbp]
\centering
\includegraphics[width=0.75\linewidth]{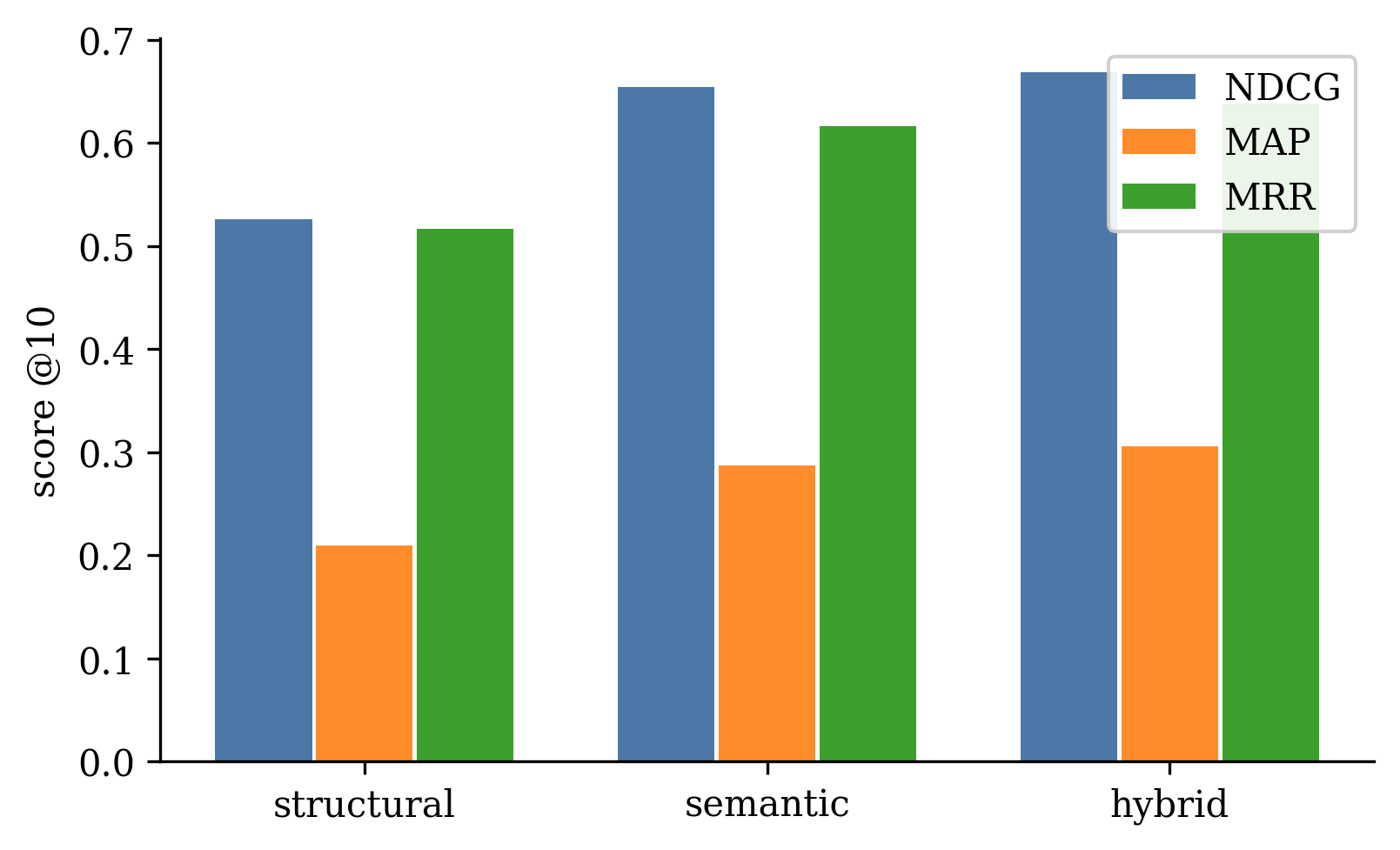}
\caption[]{Recommendation of related works: structural vs. semantic vs. hybrid method (NDCG@10, MAP, MRR; relevance = same subfield).}
\label{fig-h3-methods}
\end{figure}

We use the recommendation use case as a \textbf{controlled setting for comparing three signals}---structural, semantic, and combined. The goal is not to select a universally winning method but to show in which regimes each signal carries information. For a sample of 600 query works we rank candidates by three methods (structural = bibliographic coupling, semantic = cosine of embeddings, hybrid = late fusion with weight $\alpha$) and define relevance by membership in the same subfield; we measure NDCG@10, MAP, and MRR (Figure~\ref{fig-h3-methods}). In this use case the semantic signal (NDCG@10 = 0.654) and the fusion ($\alpha$ = 0.700, 0.668) rank more accurately than the structural signal alone (0.525) (paired permutation test, both p \textless  0.001), whereas the aggregate difference between the fusion and the semantic method alone is not significant (difference 0.014, 95\% bootstrap CI [-0.001; 0.029], p = 0.079). This is expected and fine: whether the combination is worth the higher complexity is decided by the use case and the data structure, not by a universal ranking of methods.

\begin{figure}[!htbp]
\centering
\includegraphics[width=0.75\linewidth]{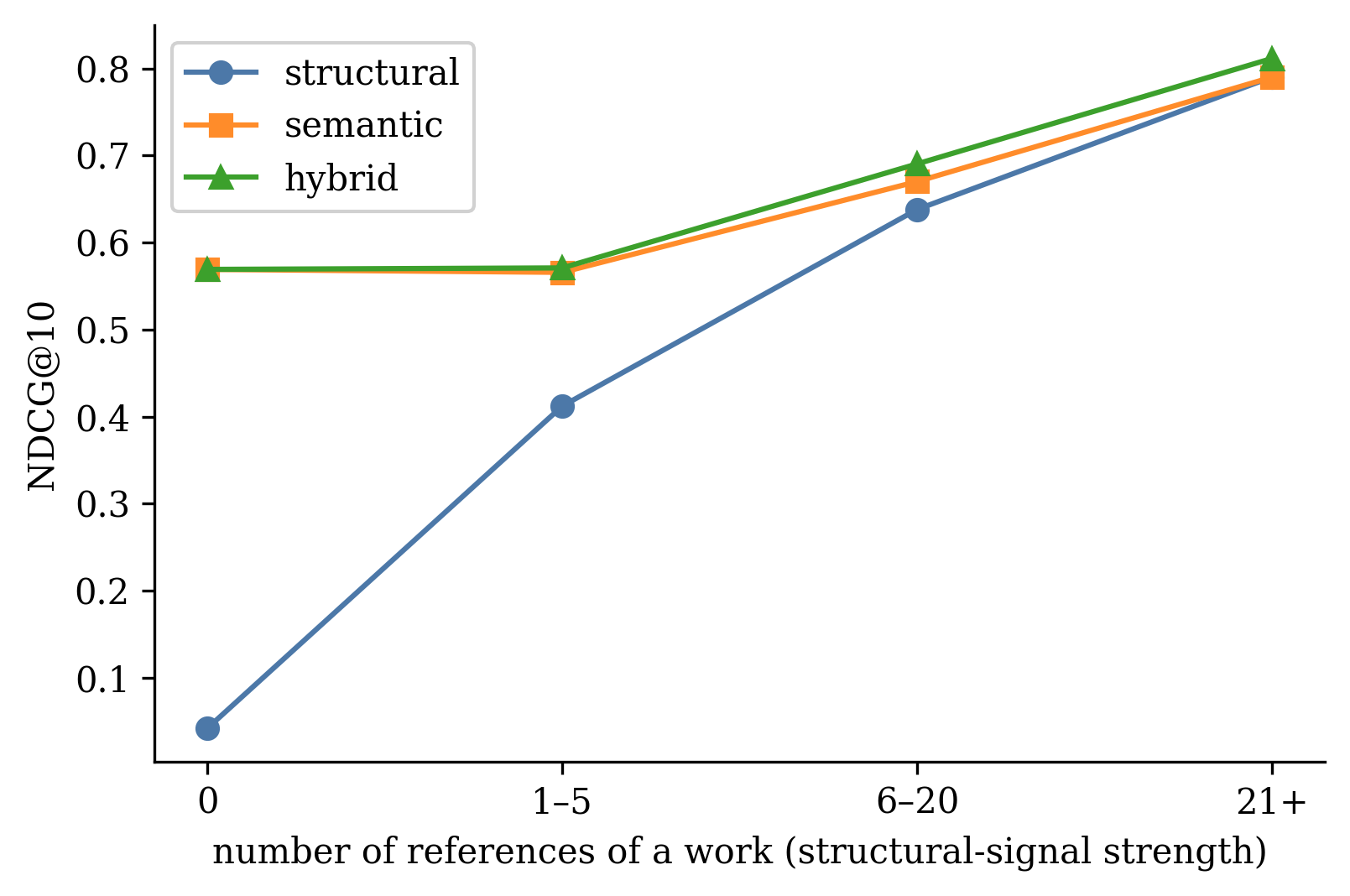}
\caption[]{NDCG@10 by the number of references of the query work (the strength of the available structural signal). Bibliographic coupling carries a weak signal for works with few references; the semantic signal is independent of the number of references. The optimal fusion weight $\alpha$ therefore depends on the task and the data.}
\label{fig-h3-coldstart}
\end{figure}

How the signals change \textbf{with the availability of structural context} is shown by stratifying the query works by their \textbf{number of references} (Figure~\ref{fig-h3-coldstart}). For works with few references, bibliographic coupling carries a weak signal (NDCG@10 = 0.042), because it has nothing to draw on, whereas the semantic signal is independent of the number of references (0.569); as the number of references grows, the structural signal improves, and for works with 21+ references it approaches the semantic one, where the combination of the two signals ranks best. This illustrates why the weight $\alpha$ is \textbf{task- and data-dependent}: the optimal ratio of structure to content depends on how much structural context is available for the given data---fusion is not an assumption of superiority but a tool for setting it.

\section{Discussion}

\textbf{What the semantic layer adds.} Embeddings capture content proximity that the graph topology does not see, and they are directly connectable to it via stable identifiers. Hybrid fusion makes it possible, for each task, to choose explicitly how much to rely on structure and how much on content; it is not an assumption that the combination always outperforms the individual layers, but an interface for steering the two complementary signals according to the nature of the task and the data.

\textbf{Limitations.} (1) \textit{Abstract coverage}: some works have no abstract in OpenAlex, and we represent such works by the title alone---the semantic signal is weaker for them. (2) \textit{Relevance via subfield}: as the reference in the recommendation use case we use thematic membership, which is an available but coarse proxy for true relevance; a stricter evaluation would use a temporal split of citations (train $\leq$ T, test T+1) and expert annotations. (3) \textit{Scale}: CPU embedding is sufficient for an institutional corpus; for national/global corpora, GPU acceleration and quantized ANN indexes are necessary. (4) \textit{Anisotropy}: contextual embeddings have a high baseline cosine similarity; what is relevant is the relative, not the absolute, ranking.

\textbf{Relation to the broader framework.} The semantic layer, together with the structural layer (the previous work) and the fusion layer, forms three components of an integrated framework for knowledge mining from the publishing network. The dissertation joins these layers and generalizes the fusion across tasks.

\section{Conclusion}

We complemented the structural graph approach with a semantic layer based on citation-informed SPECTER2 embeddings and with a modular structural--semantic fusion. On an institutional corpus we showed that the embeddings correspond to the expert thematic taxonomy better than a lexical baseline, and that in a recommendation use case the structural, semantic, and combined signals carry information in different regimes depending on the available data. We therefore formulate the fusion as a task-tunable mechanism (the weight $\alpha$ chosen according to the task), not as a universally better method. We release everything as a reproducible open-source extension of the apnet library.

Future work will lead to (1) a stricter evaluation of recommendation on temporal citation splits, (2) scaling embeddings to larger corpora with quantized ANN indexes, and (3) integrating the two layers into a unified framework and its application to anomaly detection and the identification of research gaps.

\section{Data and code availability}

The semantic and fusion layers are part of the open-source library apnet (MIT). The embeddings are computed from public OpenAlex metadata; the repository contains the exact commands to download the abstracts, compute the embeddings, and reproduce experiments H2 and H3 from a single script. The analysis requires no credentials (the OpenAlex API is open).

\bibliography{main}
\end{document}